\documentclass[11pt,a4paper]{article}
\pdfoutput=1
\usepackage{jheppub}
\usepackage{amsmath,amssymb,amsthm,amsfonts}
\allowdisplaybreaks
\usepackage{slashed} 
\usepackage{graphicx,color,float}
\usepackage{subfigure}
\usepackage{dcolumn}
\usepackage{hyperref}      
\usepackage{bm}
\usepackage{epsfig}
\usepackage{epstopdf}
\usepackage{setspace}
\usepackage{comment}
\usepackage{multirow}
\usepackage{orcidlink}

\newcommand{\be}{\begin{equation}}
\newcommand{\ee}{\end{equation}}

\usepackage[normalem]{ulem}
\usepackage{fancyvrb}
\usepackage{xcolor}

\title{Exploring the lifetime frontier with a beam-dump experiment at CiADS}

\author[a,b,c]{Liangwen Chen,\,\orcidlink{0000-0002-1592-288X}}
\emailAdd{chenlw@impcas.ac.cn}

\author[d,e]{Mingxuan Du,\,\orcidlink{0000-0002-1938-5794}}
\emailAdd{mingxuandu@pku.edu.cn}

\author[a,b,c,1]{Zhiyu Sun,\,
\note{Corresponding author: \href{mailto:sunzhy@impcas.ac.cn}{sunzhy@impcas.ac.cn}}}
\emailAdd{sunzhy@impcas.ac.cn}

\author[f]{Zeren Simon Wang,\,\orcidlink{0000-0002-1483-6314}}
\emailAdd{wzs@hfut.edu.cn}

\author[g]{Fang Xie,\,\orcidlink{0000-0003-3854-7231}}
\emailAdd{xiefang@fudan.edu.cn}

\author[a,c,h]{Ju-Jun Xie,\,}
\emailAdd{xiejujun@impcas.ac.cn}

\author[a,b,c]{Lei Yang,\,}
\emailAdd{lyang@impcas.ac.cn}

\author[a,b]{Pei Yu,\,}
\emailAdd{yupei@gdlhz.ac.cn}

\author[f]{Yu Zhang\,\orcidlink{0000-0001-9415-8252}}
\emailAdd{dayu@hfut.edu.cn}

\affiliation[a]{Heavy Ion Science and Technology Key Laboratory, Institute of Modern Physics, Chinese Academy of Sciences, Lanzhou 730000, China}
\affiliation[b]{Advanced Energy Science and Technology Guangdong Laboratory, Huizhou 516000, China}
\affiliation[c]{School of Nuclear Science and Technology, University of Chinese Academy of Sciences, Beijing 101408, China}
\affiliation[d]{School of Physics and State Key Laboratory of Nuclear Physics and Technology, Peking University, Beijing 100871, China}
\affiliation[e]{Center for High Energy Physics, Peking University, Beijing 100871, China}
\affiliation[f]{School of Physics, Hefei University of Technology, Hefei 230601, China}
\affiliation[g]{Institute of Modern Physics, Fudan University, Shanghai 200433, China}
\affiliation[h]{Southern Center for Nuclear-Science Theory (SCNT), Institute of Modern Physics, Chinese Academy of Sciences, Huizhou 516000, Guangdong, China}

\date{\today}

\abstract{We propose a beam-dump experiment (BDE) at the upcoming facility of China initiative Accelerator Driven System (CiADS), called CiADS-BDE, in order to search for long-lived particles (LLPs) predicted in various beyond-the-Standard-Model (BSM) theories. The experiment is to be located in the forward direction of the incoming low-energy proton beam at CiADS, leveraging the strong forward boost of the produced particles at the beam dump in general. The space between the dump and the detector is largely available, allowing for installation of shielding and veto materials and hence low levels of background events. We elaborate on the detector setup, and choose dark photon as a benchmark model for sensitivity study. We restrict ourselves to the signature of an electron-positron pair and perform detailed background estimates. We find that with 5 years' operation, unique, currently unexcluded parts of the parameter space for $\mathcal{O}(100)$~MeV dark-photon masses and $\mathcal{O}(10^{-9}\text{--}10^{-8})$ kinetic mixing can be probed at the CiADS-BDE. Furthermore, considering that there is no need to set up a proton beam specifically for this experiment and that the detector system requires minimal instrumentation, the experiment is supposed to be relatively cost-effective. Therefore, we intend this work to promote studies on the sensitivity reach of the proposed experiment to additional LLP scenarios, and in the end, the realization of the experiment. Incidentally, we study the sensitivity of the same BDE setups at the High Intensity Heavy-ion Accelerator Facility (HIAF), presently in operation near the CiADS program site, and conclude that HIAF-BDE could probe new parameter regions, too.}

\begin{document}
\maketitle

\section{Introduction}\label{sec:intro}

In nuclear waste exist long-lived, radioactive isotopes, posing potential risk to the environment if not properly treated.
Currently, the optimal technique for their transmutation into short-lived isotopes is Accelerator Driven System (ADS).
In Huizhou, Guangdong, China, the facility, China initiative Accelerator Driven System (CiADS)~\cite{liu2017physics,Liu:2019cdo,Wang:2019ddc,Liu:2019xgd,He:2023izb,Cai:2023caf,Wang:2024joa}, is presently under construction and slated to start operation in 2028.
It will be the first prototype ADS facility at megawatt level for studying and developing relevant technologies of nuclear waste disposal in the world.
Its scientific goals include improving the long-term performance of a superconducting linear accelerator, high-power spallation target, and sub-critical reactor core or cladding.
Particularly, the accelerator-driven subcritical reactor imposes extremely high requirements on the stability of the accelerator beam.
The superconducting linear accelerator of CiADS needs to conduct long-term stability experiments with high-power beam operation, which requires a high-power beam dump for beam commissioning.
It is designed based on a 600~MeV proton-beam energy\footnote{In the context of ADS facilities, the energy of a proton beam usually refers to the kinetic energy of the protons.} and a current of 0.5~mA, with a 3~MW upgrade plan (with proton energy of 600~MeV and current of 5~mA) currently under consideration.
A further upgrade with proton energy of 2~GeV and a current of 5~mA is also expected.
The target of CiADS is supposed to consist of oxygen-free copper.

During the beam commissioning, the beam dump is bombarded with the proton and large numbers of protons on target (POT) are expected. As a result there should be copious production of light mesons at the target that travel with a large boost in the forward direction of the incoming proton beam in the laboratory frame, including pions and $\eta$ mesons depending on the proton-beam energy.
These light mesons could undergo rare decays into new, light, fundamental particles predicted in beyond-the-Standard-Model (BSM) theories, which then travel in roughly the same direction.
Moreover, such light particles can be produced in either direct proton-proton collisions or electromagnetic cascades inside the beam dump.
These proposed new particles are often predicted to be long-lived particles (LLPs) for either their small mass, weak couplings with other particles, a heavy mediator responsible for their decays, or suppressed phase space.
Once produced, they travel a distance ranging from a few mm to several kilometers, or even longer, before decaying into BSM or Standard-Model (SM) particles.
Indeed, in recent years, LLPs have become one of the major directions in searches for BSM physics, considering the fact that so far at the LHC no new heavy fundamental particles have been discovered.
See Refs.~\cite{Alimena:2019zri,Lee:2018pag,Curtin:2018mvb,Beacham:2019nyx} for recent reviews of the LLP models and experimental searches.

We propose a beam-dump experiment (BDE) at the CiADS facility, called CiADS-BDE, in order to search for LLPs produced at the beam dump.
The experiment should be implemented as a detector with tracking capabilities to allow for the reconstruction of a displaced vertex (DV) consisting of charged particles that emerge from the LLP decays.
Similar ideas have been proposed at CERN, Switzerland, such as FASER~\cite{Feng:2017uoz,FASER:2018eoc}, FACET~\cite{Cerci:2021nlb}, and SHiP~\cite{SHiP:2015vad,Alekhin:2015byh,SHiP:2021nfo,Albanese:2878604}, all essentially making use of the strong forward boost possessed by particles arising from proton-proton collisions.
Compared to these experiments, the CiADS-BDE has a proton beam of lower energies, disallowing its sensitivity reach to higher LLP masses; however, its high intensity is expected to provide strong sensitivities in the low-mass regime.
The detector can be placed a few meters behind the dump, and the space between the detector and the dump is available for implementation of veto or shielding materials such as lead for background suppression.

In this work, we choose light dark photons~\cite{Okun:1982xi,Galison:1983pa,Holdom:1985ag,Boehm:2003hm,Pospelov:2008zw} as a representative LLP scenario, to illustrate the overall performance of CiADS-BDE for such searches.
The dark photons arise naturally when a new U(1) gauge group is appended to the SM gauge group, of which the gauge boson mixes kinetically with the SM U(1) gauge boson, with a term $-\frac{\epsilon}{2}F^{\mu\nu}F^{\prime}_{\mu\nu}$; here, $\epsilon$ labels the kinetic-mixing strength, and $F^{\mu\nu}$ and $F^{\prime}_{\mu\nu}$ are the field strength tensors of the SM and the new U(1) gauge bosons, respectively.
The dark photon can be a mediator particle bridging the visible sector (our SM spectrum) and a hidden sector consisting of dark matter (DM) and potentially other particles.
If discovered, it could elucidate the nature of the DM.
In addition, the dark photon in the minimal scenario can be constrained by various experiments.
For these reasons among others, searches for dark photons have been conducted that are either promptly decaying or long-lived, setting bounds on the dark-photon kinetic mixing for masses above 1 MeV in particular.
Collider searches focus on signatures of dark-photon decays at or slightly displaced from the interaction point of the incoming beams~\cite{CMS:2018rdr,CMS:2018lqx,ATLAS:2015itk,ATLAS:2014fzk,ATLAS:2016jza,LHCb:2017trq,BaBar:2014zli,LHCb:2019vmc}, and fixed-target and beam-dump experiments have searched for largely displaced decays of the dark photons~\cite{Riordan:1987aw,Bross:1989mp,Davier:1989wz,NA482:2015wmo,Bjorken:1988as,Batell:2014mga,Marsicano:2018krp,Blumlein:2011mv,Blumlein:2013cua,Gninenko:2012eq}.
For masses below $\mathcal{O}(100)$~MeV, bounds are also derived from the constraints on energy losses in supernovae~\cite{Dent:2012mx,Dreiner:2013mua,Chang:2016ntp,Hardy:2016kme,Bottaro:2023gep}.
Furthermore, the precise measurements of the electron magnetic moments have been used for setting indirect bounds~\cite{Pospelov:2008zw}.
Additionally, cosmic microwave background and nucleosynthesis~\cite{Fradette:2014sza} have been considered for obtaining constraints for dark-photon mass between 1~MeV and 100~MeV.
We refer to Ref.~\cite{Ilten:2018crw} for a summary of both existing and future bounds on the dark photons.

Before closing the section, we mention another program, High Intensity Heavy-ion Accelerator Facility (HIAF)~\cite{Yang:2013yeb,Zhou:2022pxl,Yang:2023hfx,Chen:2024wad,Liu:2024pgf}, located in the vicinity of the CiADS, currently in operation.
It is one of the leading heavy-ion accelerators in the world.
At this heavy-ion research facility, proton beams of energy up to 9.3~GeV will hit heavy-metal targets (which we assume to be identical to that used at the CiADS experiment, i.e.~oxygen-free copper), thus involving physics similar to that discussed above.
However, despite the larger proton-beam energy, its duty factor is assumed to be much smaller than that of the CiADS, and as a result, around three orders of magnitude fewer POT per year are planned.
Nevertheless, we take into account the possibility of setting up a similar BDE at HIAF (HIAF-BDE) and perform sensitivity studies incidentally.

This paper is organized as follows.
In Sec.~\ref{sec:experiment} we discuss the setup of the CiADS-BDE, including possible locations and requirements on the detector.
We then introduce in Sec.~\ref{sec:model} our benchmark model, the dark photon, including its various production and decay modes, as well as relevant background sources and their estimates.
Sec.~\ref{sec:production_signal} is devoted to discussions on possible production modes of the dark photons and computation procedures of the signal-event numbers.
The numerical results on the sensitivity reach of the CiADS-BDE to the dark photon are presented in Sec.~\ref{sec:results}.
Finally, we conclude the paper and provide an outlook in Sec.~\ref{sec:conclusions}.

\section{Proposed beam-dump experimental setups}\label{sec:experiment}

\setlength{\extrarowheight}{4pt}
\begin{table}[t]
\centering
\begin{tabular}{|c|c|c|c|}
\hline
Facility & Proton Energy @ Current  &  Duty factor & POT per year      \\
\hline
\multirow{3}{*}{CiADS} &
500/600 MeV @ 0.5~mA  &  $75\%$ & $6.6\times 10^{22}$      \\
\cline{2-4}
&500/600 MeV @ 5~mA  &  $75\%$ & $6.6\times 10^{23}$      \\
\cline{2-4}
& 2~GeV @ 5~mA &  $75\%$ & $6.6\times 10^{23}$      \\
\hline
HIAF  & 9.3~GeV @ 0.024~mA &  $8.3\%$   &   $3.9\times 10^{20} $ \\
\hline
\end{tabular}
\caption{Beam parameters, assumed duty factors, and the corresponding POT per year, at CiADS and HIAF.}
\label{tab:beam_pot}
\end{table}

In Table~\ref{tab:beam_pot}, we list the beam energy, beam current, duty factor, and the corresponding numbers of POT per year for the beam commissioning at CiADS and HIAF.
For deriving the duty factor, we assume 9-month (1-month) beam commissioning time per year for CiADS (HIAF).
For the relatively low-energy beam setups of the CiADS, we will focus on the proton kinetic-energy of 600 MeV with $6.6\times 10^{23}$~POT per year, for numerical studies.
In the sensitivity analysis, as a benchmark we will take 5 years' operation duration for the proton beams at both CiADS and HIAF.

\begin{figure}[t]
 \centering
 \includegraphics[width=0.7\textwidth]{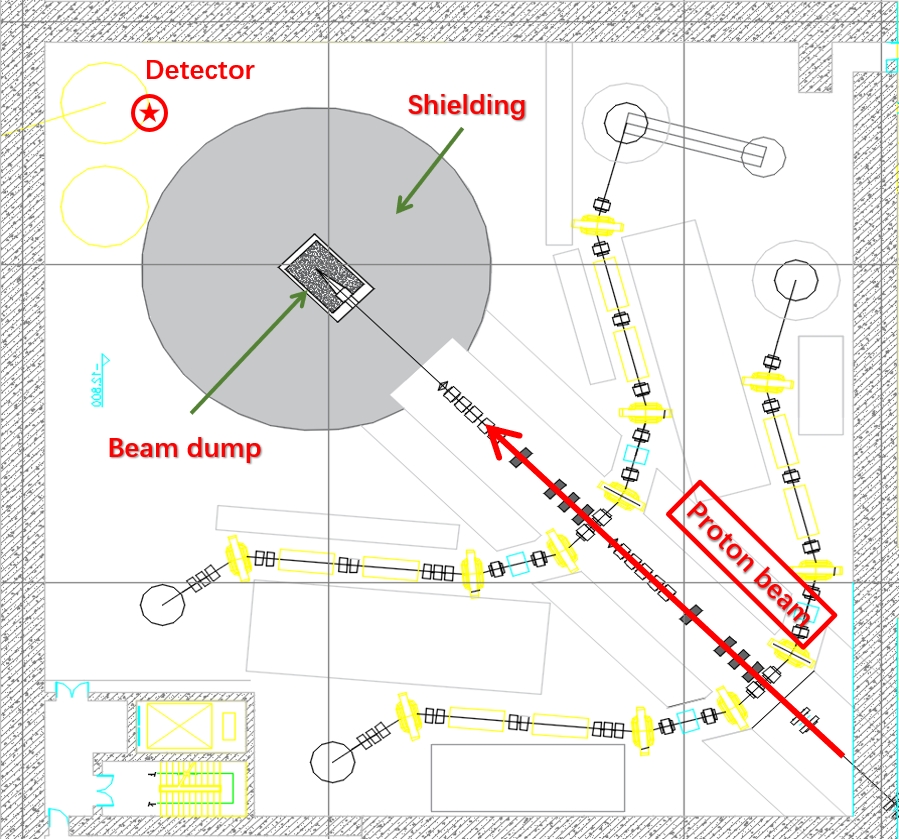}
 \caption{Preliminary layout of the CiADS-BDE.
 The red circle with a star inside denotes the potential location of the detector of the CiADS-BDE, with a distance of 10 meters from the dump.
 }
 \label{fig:exp_setup}
\end{figure}

In Fig.~\ref{fig:exp_setup}, we show a preliminary layout of the proposed CiADS-BDE.
Regarding the high-power beam dump, we propose to use oxygen-free copper as the main material, because it meets the requirements for high reliability and cost-effectiveness.
Additionally, the high thermal conductivity of copper is beneficial for removing the thermal power arising from the bombardment.
A segmented insertion design allows for reasonable power distribution and independent replacement.
The CiADS beam-dump design thus accommodates both direct collection and post-target collection needs.
With a beam power of several MW, the thickness of concrete material on the beam-dump station is required to be about 7 meters, in order to shield radiation.

As discussed in Sec.~\ref{sec:intro}, LLPs can arise from different production mechanisms.
For the dark-photon scenario on which we focus, we will restrict ourselves to the signal decays of the dark photon into an electron-positron pair.
We note that the kinematics of the electron and positron inside the detector depend mainly on the dark-photon mass and the production mode.
In this proposal, we adopt a baseline design similar to that of the SHiNESS detector proposed at the European Spallation Station~\cite{Soleti:2023hlr}, that is, a cylindrical liquid-scintillator detector.
We further propose to place the detector behind the dump with aligned orientation, in order to leverage the forward boost of the produced LLPs for the purpose of achieving better acceptance rates.
The fiducial volume is supposed to have a radius of 0.1~m or 1~m, and a length of 1~meter.
It should be filled with liquid scintillator loaded with gadolinium.
With detector-level simulations, we require that the reconstructed energy of both the electron and the positron should be larger than 17 MeV and their opening angle should be at least $30^\circ$, for the purpose of suppressing background events.
Further, in the baseline design Incom Large Area Picosecond Photodetectors (LAPPDs{$^{\mathrm{TM}}$}) are employed to enable the discrimination between the Cherenkov and scintillation light emission, which can improve the vertex resolution and allow the reconstruction of the directionality of charged particles including electrons and positrons.

The proposed location of the detector is displayed in Fig.~\ref{fig:exp_setup} as a red circle with a star inside.
The detector is to be placed 10 m behind the beam dump.
If a shorter distance is taken, we expect larger acceptance rates of the detector to the LLPs but also higher levels of background events.
For the HIAF-BDE, we assume similar designs and locations of the detector behind the dump.

Finally, we comment that this proposed beam-dump experiment is supposed to be particularly economical because the produced LLPs are by-products of the beam commissioning and other applications of the beam, and only minimal instrumentation of the detector systems is required.

\section{The dark-photon model}\label{sec:model}

The low-energy interaction Lagrangian of the dark photon with the SM particles is given as follows,
\begin{eqnarray}
    \mathcal{L}_{\text{int.}}=-\epsilon e J^\mu \gamma^\prime_{\mu},
\end{eqnarray}
with $\epsilon$ denoting the kinetic-mixing angle and $J^\mu$ labeling the SM electromagnetic current.
Via this interaction term, the dark photon can be produced at the CiADS-BDE in various modes~\cite{Kyselov:2024dmi}, including rare meson decays and proton bremsstrahlung,\footnote{In principle, the dark photons can also arise via mixing with vector mesons such as $\rho^0$ and $\omega$, in Drell-Yan processes, and from interaction of secondary particles inside the beam dump~\cite{Blinov:2024pza,Zhou:2024aeu}. The contributions of the former possibility are tiny, if not negligible, considering the relatively low beam energies at CiADS. The same conclusion holds for the Drell-Yan processes. The latter option should only give contributions for dark-photon mass below 100 MeV and the corresponding parameter regions have been mostly excluded by past beam-dump experiments and astrophysical observations; see the relevant discussion in, e.g.~Refs.~\cite{Gorbunov:2014wqa,Kyselov:2024dmi,Zhou:2024aeu}.} and then decay into SM particles.
Below, we discuss them separately.

\subsection{Dark-photon production}\label{subsec:production}

\subsubsection{Meson decays}\label{subsubsec:meson_decays}

Light mesons can be copiously produced at the CiADS-BDE, provided the large numbers of POT expected.
Their decays can be the dominant source of the dark photons at the facility.
Since these light mesons have a strong forward boost and decay promptly, the dark photons produced therefrom inherit the kinematics and travel towards the detector behind the CiADS-BDE, leading to excellent acceptance rates.

Here, we focus on the rare decays of the $\pi^0$ and $\eta$ mesons into a SM photon and a dark photon, which make the primary contributions.
The corresponding decay branching ratios can be computed with the following expressions~\cite{Batell:2009di},
\begin{eqnarray}
    \text{BR}(\pi^0\to \gamma \gamma') &=& 2\epsilon^2 \, \Big( 1 - \frac{m^2_{\gamma'}}{m^2_{\pi^0}}   \Big)^3 \, \text{BR}(\pi^0\to \gamma\gamma),\\
    \text{BR}(\eta\to \gamma \gamma') &=& 2\epsilon^2 \, \Big( 1 - \frac{m^2_{\gamma'}}{m^2_{\eta}}   \Big)^3 \, \text{BR}(\eta\to \gamma\gamma),
\end{eqnarray}
where $m_{\pi^0}=134.98$~MeV and $m_{\eta}=546.86$~MeV are respectively the masses of the $\pi^0$ and $\eta$ mesons~\cite{ParticleDataGroup:2024cfk}.
Further, BR$(\pi^0/\eta \to \gamma\gamma)$ is the decay branching ratio of the $\pi^0/\eta$ meson into a pair of the SM photons, taken to be 99\% and 39\%, respectively~\cite{ParticleDataGroup:2024cfk}.

\subsubsection{Proton bremsstrahlung}\label{subsubsec:model_pb}

Another important source of forward-going dark photons is proton bremsstrahlung (PB) in $pp$ collisions, namely $pp\to \gamma' X$, with $p$ and $X$ denoting a proton and anything, respectively.
The PB process can give competitive contributions to dark-photon production compared to meson-decay (MD) processes~\cite{Blumlein:2013cua, deNiverville:2016rqh, Tsai:2019buq,  Feng:2017uoz, Foroughi-Abari:2021zbm, Du:2021cmt, Du:2022hms, Du:2023hsv}.

At the CiADS-BDE and HIAF-BDE, we consider the $p Cu \to \gamma' X$ process for which we will use $p_p$, $p_{Cu}$, and $k$ to represent the momenta of the proton, copper nucleus, and dark photon in the lab frame, respectively.
The corresponding differential cross section ${d^2 \sigma_{\rm PB}}/{d E_{k} d\cos\theta_{k}}$ is given by~\cite{Du:2022hms}
\be
\frac{d^2 \sigma_{\rm PB}}{d E_{k} d\cos\theta_{k}} =
\left|\frac{\vec{k}}{\vec{k}^0}\right| \left| F_V (k) \right|^2
\left|F_* \left(p_p-k\right)\right|^2
\frac{d^2 \mathcal{P}_{p \to \gamma' p}}{d E^0_{k} d \cos\theta_{k}^0}
\sigma_{pCu}(s'),
\label{eq:d2sigma: lab:frame}
\ee
where $(E^0_{k}, \vec{k}^0)$ and $\theta_{k}^0$, and $(E_{k}, \vec{k})$ and $\theta_{k}$, represent the four-momentum and polar angle for the dark photon in the center-of-mass (CM) and laboratory frames, respectively.
The term $|\vec{k}|/|\vec{k}^0|$ is the Jacobian between the CM-frame variables $(E^0_k, \cos\theta^0_k)$ and the lab-frame variables $(E_k, \cos\theta_k)$.
$F_V$ denotes the vector-meson form factor, accounting for the $p \gamma' p$ vertex, and $F_*$ labels the off-shell form factor, which accounts for the off-shell-ness of the proton after bremsstrahlung.
$d^2 \mathcal{P}_{p \to \gamma' p}/d E_k^0 d \cos\theta_{k}^0$ is called the splitting kernel and its analytic form will be given below.
Finally, $\sigma_{pCu}$ is the proton-copper cross section evaluated at $s'=\left(p_p + p_{Cu} - k \right)^2$.
Since the proton-copper cross section is of the order of $O(100)$~mb within the energy range of interest, for simplicity, we take $ \sigma_{pCu} (s') = \sigma_{pCu}$.
Consequently, given the known $N_{\rm POT}$, the signal number for the PB process, as computed per Eq.~\eqref{eqn:NS_PB} to be given below, is independent of $\sigma_{pCu}$.

According to the vector-meson-dominance model which assumes that a photon interacts with a proton via exchanging vector mesons,
$F_V$ is given by~\cite{Faessler:2009tn} 
\be
F_V(k) = \sum_{V =\rho \, \rho' \, \rho'' \, \omega \, \omega' \, \omega''} \frac{f_V m_V^2}{m_V^2 - k^2 - i m_V \Gamma_V}, 
\label{eq:PB_formfactor}
\ee
where $m_V$ ($\Gamma_V$) is the mass (total decay width) of the vector meson $V$, $m_{\rho} = m_{\omega} = 0.77$~GeV, $m_{\rho'} = m_{\omega'} = 1.25$~GeV, $m_{\rho''} = m_{\omega''} = 1.45$~GeV, $\Gamma_{\rho} = 0.150$~GeV,
$\Gamma_{\omega} = 0.0085$~GeV, $\Gamma_{\rho'} = \Gamma_{\omega'} = 0.3$~GeV, $\Gamma_{\rho''} = \Gamma_{\omega''} = 0.5$~GeV, $f_{\rho} = 0.616$, $f_{\rho'} = 0.223$, $f_{\rho''} = -0.339$, $f_{\omega} = 1.011$, $f_{\omega'} = -0.881$, and $f_{\omega''} = 0.369$.
The off-shell form factor $F_*$ is calculated with~\cite{Feuster:1998cj, Foroughi-Abari:2021zbm}
\be
F_* (p') = \frac{\Lambda^4}{ \Lambda^4 + (p'^2 - m_p^2)^2},
\label{eq: off-shell FF}
\ee
where $m_p$ is the proton mass, $p'$ is the four-momentum of the proton after bremsstrahlung, and $\Lambda$ is a hard cut-off scale which we fix at 1.5~GeV following the practice taken in Ref.~\cite{Foroughi-Abari:2021zbm}.

We do not employ the traditional splitting kernel computed with the Fermi-Weizs\"acker-Williams approximation~\cite{Fermi:1924, Williams:1934, Weizsacker:1934}, as it requires relativistic and collinear conditions for both initial- and final-state particles which are not met in low-energy proton beam-dump experiments. 
Instead, we adopt the method described in Ref.~\cite{Du:2022hms} for calculating the PB process, 
which is applicable to proton beams of any energy and full angles between the final and initial states.
The splitting kernel is given by~\cite{Du:2022hms} 
\begin{equation}
\frac{d^2 \mathcal{P}_{p \to \gamma' p}}{d E_k^0 d \cos\theta_{k}^0}  
= \frac{2 \epsilon^2 \alpha}{\pi E_k^0} 
 \frac{2\beta_f \beta_k}{(3-\beta_f^2)\beta_i}
\frac{N}{s\left[x^2-(1-y)^2\right]^2},
\label{eq:SK:analytic}
\end{equation}
where 
$\beta_i=\sqrt{1-4m_p^2/s}$, $\beta_f=\sqrt{1-4m_p^2/s_k}$ with $s_k=s+m_{\gamma'}^2-2 E_k^0 \sqrt{s}$, $\beta_k=\sqrt{1-m_{\gamma'}^2/(E_k^0)^2}$, $x=\beta_i \beta_k \cos\theta_{k}^0$, $y= m_{\gamma'}^2/(E_k^0\sqrt{s})$, and 
\begin{align}
N = &  
(1-y)^2
\left[ \beta_i^2 (2 m_p^2+s_k+m_{\gamma'}^2) + 
(E_k^0)^2 \beta_k^2 \right]
\nonumber \\ & 
- \left[ 2 m_p^2  + m_{\gamma'}^2(1+s_k/s)+ s_k \right]x^2 
- (E_k^0)^2 x^4.
\end{align}

We also note that while our analysis of dark-photon production via the PB process employs vector-meson form factors that involve vector mesons only, our computation of dark-photon production in meson decays considers solely the contributions from light pseudoscalar mesons $\pi^0$ and $\eta$, excluding the vector mesons. 
Therefore, our approach avoids any potential double counting between the PB and MD processes.

\subsection{Dark-photon decay}\label{subsec:decay}

The dark photon decays via the kinetic mixing into a pair of leptons or quarks at the parton level.
To compute the partial decay widths into leptonic and hadronic final states, we make use of the formulas given below~\cite{Bauer:2018onh,Batell:2009yf,Fabbrichesi:2020wbt},
\begin{eqnarray}
    \Gamma(\gamma'\to \ell^- \ell^+)&=&\frac{1}{3}\alpha_{\text{QED}} \epsilon^2 m_{\gamma'}\sqrt{1-\frac{4m^2_{\ell}}{m^2_{\gamma'}}}\Big(  1 + \frac{2m^2_{\ell}}{m^2_{\gamma'}} \Big),\label{eqn:width_dp_ll}\\
    \Gamma(\gamma' \to \text{hadrons})&=& \Gamma(\gamma'\to \mu^- \mu^+)\times R(s=m^2_{\gamma'}),\label{eqn:width_dp_hadrons}
\end{eqnarray}
where $\alpha_{\text{QED}}=1/137$ is the fine structure constant, $\ell=e,\mu,\tau$ labels an SM charged lepton, and $R(s)=\sigma_{e^-e^+\to \text{hadrons}}/\sigma_{e^-e^+\to \mu^-\mu^+}$ is a ratio of scattering cross sections extracted from Ref.~\cite{Ilten:2018crw}.
To compute the total width of the dark photon $\Gamma_{\gamma'}$, we sum Eq.~\eqref{eqn:width_dp_ll} and Eq.~\eqref{eqn:width_dp_hadrons}, taking into account all kinematically allowed decay channels including $e^-e^+$, $\mu^- \mu^+$, and $\pi^- \pi^+$.

\begin{figure}[t]
	\centering
	\includegraphics[width=0.495\textwidth]{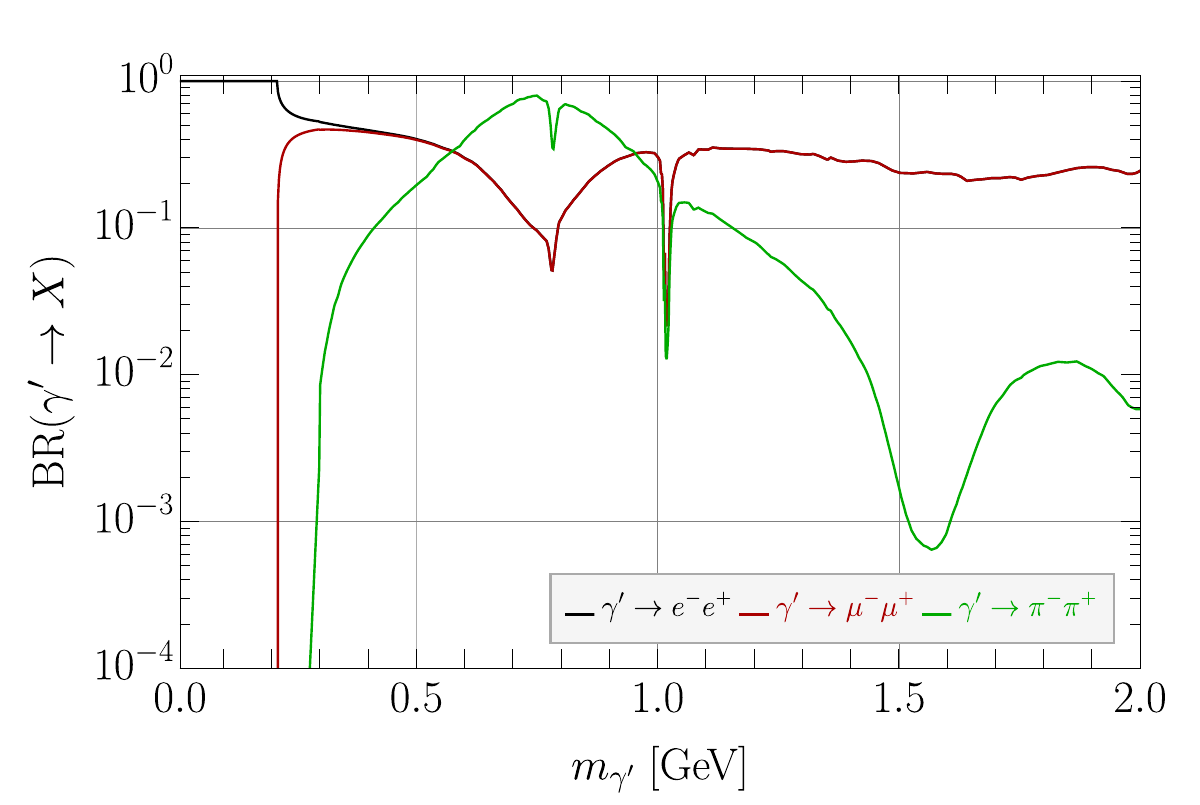}
	\includegraphics[width=0.495\textwidth]{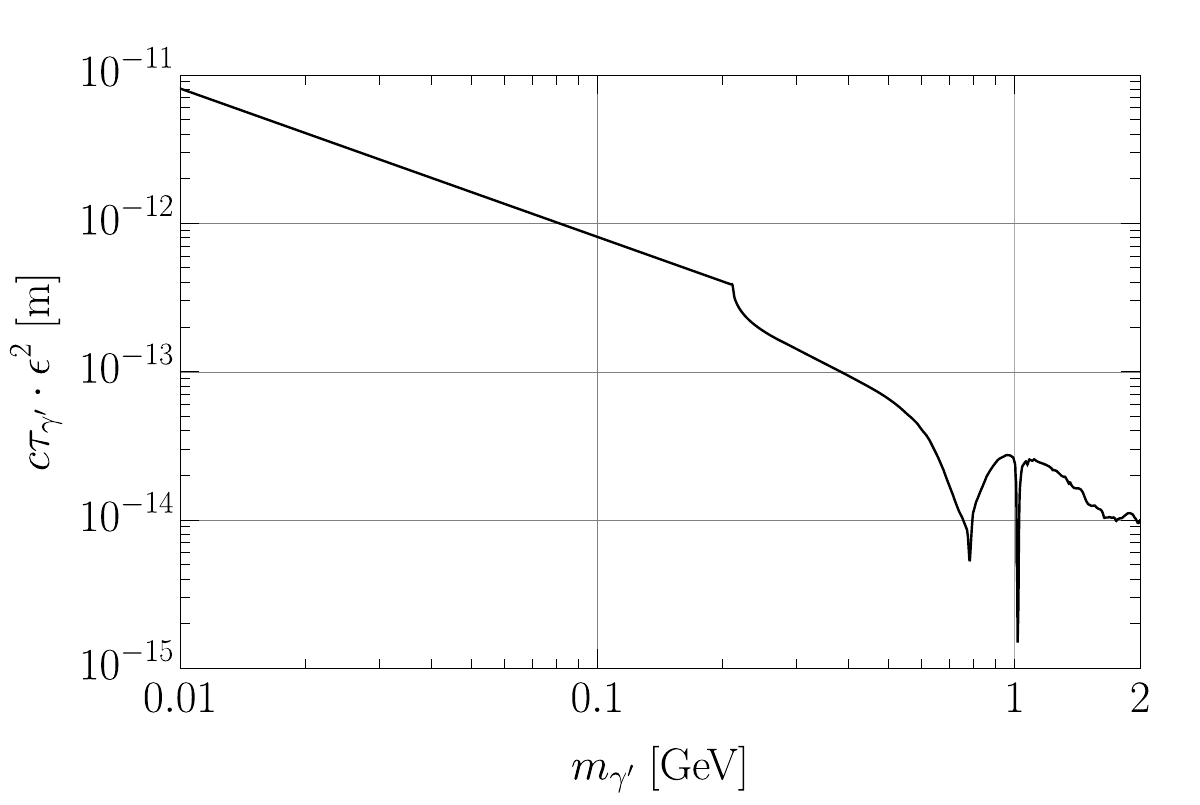}
	\caption{Left panel: decay branching ratios of the dark photon as functions of its mass, and right panel: proper decay length of the dark photon scaled with $\epsilon^2$ as a function of its mass. In the left plot, for $m_{\gamma'}\gtrsim 0.4$~GeV, the curves of the electron-pair and muon-pair channels converge.
 }
 \label{fig:dp_decay}
\end{figure}
In Fig.~\ref{fig:dp_decay}, we display two plots for BR$(\gamma'\to X)$ and $c\tau_{\gamma'}\cdot \epsilon^2$, respectively, as functions of $m_{\gamma'}$, where $X$ denotes anything and $c\tau_{\gamma'}$ is the proper decay length of the dark photon.

\subsection{Background discussion}
\label{sec:background}

The dominant background arises from misreconstruction events that can mimic a dark-photon signal, characterized by two leptons originating from the same vertex.
There are four major background sources: cosmic-ray interactions, neutrino charged-current (CC) interactions, neutrino neutral-current (NC) interactions, and the intrinsic beam component of electron antineutrinos $(\bar{\nu}_e)$.

To estimate these backgrounds, we employ a simplified simulation approach using GEANT4~\cite{GEANT4:2002zbu}.
Instead of tracking the initial neutrinos, the simulations start from the secondary particles, and propagate them through the detector geometry.
The detector response was modeled according to a typical liquid-scintillator detector, assuming an energy resolution~$\sim $10\% and a timing resolution of the order of nanoseconds.
For the purpose of a conservative estimate, we take the larger detector setup, with a radius of 1~m, for the background modeling.

The simulation tracks the energy deposits in the liquid scintillator.
Clustering algorithm is then applied to form clusters, from which event properties including the vertex, time, and energy, are reconstructed. 
Prior to further analysis, a dedicated cut was applied to the reconstructed events: two leptons are required to have an energy larger than 17~MeV and an opening angle larger than $30^\circ$.

\subsubsection{Cosmic-Ray background}
\label{subsec:bkg_cosmic}

The cosmic-ray background consists of two components: direct cosmic muons and muon-induced spallation isotopes.

The direct cosmic high-energy muons that traverse the detector can be misreconstructed as a pair of charged leptons.
We propose a muon veto system with plastic scintillator panels to mitigate this background, expected to reduce the residual rate to below 1 event per year.

Cosmic muons interacting with the detector materials and the surrounding shielding can induce spallation reactions, producing short-lived radioactive isotopes such as $^9\mathrm{Li}$ and $^8\mathrm{He}$.
The subsequent beta decays of these isotopes can produce events that mimic signal-like events.
These decay products are not associated with a primary muon event and thus cannot be rejected by the veto system.
However, the energies of the decay products are predominantly below 17 MeV and therefore the energy cut can eject essentially all of these low-energy spallation events.

As a result,the combined use of the muon veto and the energy threshold ensures the total cosmic-ray backgrounds should be kept at a negligible level.

\subsubsection{Background from neutrino charged-current interactions}
\label{subsec:bkg_cc}

A relevant beam-related background arises from the CC interaction of $\nu_e$ with $^{12}\mathrm{C}$ in the scintillator via the process $^{12}\mathrm{C} + \nu_e \rightarrow \, ^{12}\mathrm{N}^* + e^-$.
The excited $^{12}\mathrm{N}^*$ nucleus de-excites via prompt proton emission to $^{11}\mathrm{C}$, accompanied by gamma-ray emission in the 2-7~MeV range~\cite{LSND:2001fbw}.
If the electron from the initial interaction is misreconstructed as two separate leptons, and if the subsequent low-energy gamma cascade or the delayed beta decay is not identified, this process could mimic a signal event.
With the 17~MeV energy cut this background is estimated to be less than 1~event per year.  

\subsubsection{Background from neutrino neutral-current interactions}
\label{subsec:bkg_nc}

Another beam-related background originates from the NC interactions of three neutrinos: $\nu_e$, $\nu_\mu$, and $\bar{\nu}_\mu$.
The dominant process is the exclusive excitation $^{12}\mathrm{C} + \nu_x \rightarrow \, ^{12}\mathrm{C}^* + \nu_x'$ which produce a 15.11~MeV prompt gamma.
The event can be misreconstructed as two leptons; however, with the 17~MeV energy cut, the contributions from the NC interaction can be effectively eliminated.

\subsubsection{Background from the intrinsic $\bar{\nu}_e$ beam component}
\label{subsec:bkg_ibd}

The beam dump inevitably produces a small intrinsic flux of $\bar{\nu}_e$ from $\mu^-$ decays in flight.
The primary interaction channel for this component in the liquid scintillator is inverse beta decay: $\bar{\nu}_e + p \rightarrow e^+ + n$, where a prompt positron signal is produced, followed by a delayed neutron capture.
If the delayed neutron is lost and the positron is misreconstructed as two separate leptons, the event becomes a background.
This contribution is estimated to negligible.

In the end, we conclude essentially vanishing background levels at the CiADS-BDE with a radius of 1~m are expected, and even less backgrounds at the detector with a smaller radius.

\section{Production modes and signal-event numbers}\label{sec:production_signal}

We focus on lepton-pair signatures in this work.
Specifically, we look for $e^- e^+$ pairs, which is a typical signature predicted to arise from various LLP candidates including the dark photon.
As shown in the left plot of Fig.~\ref{fig:dp_decay}, this decay mode is dominant for $m_{\gamma'}\lesssim 0.56$~GeV, overtaken by the di-pion mode for larger masses.
In addition, we impose the requirements that both electron and positron should have an energy larger than 17 MeV and their opening angle should be wider than $30^\circ$.
These kinematic cuts are implemented in order to reduce greatly the background levels and to accommodate the angular resolutions expected for the proposed detector; see Sec.~\ref{sec:experiment} for more detail.

To compute the signal-event rates for the different production modes, we apply different computation or simulation procedures.
The detail is given below for each production mode separately.

\subsection{Meson decays}\label{subsec:NS_mesondecay}

The signal-event numbers $N_S^{\pi^0/\eta}$ for dark photons from meson decays are computed with the following formula,
\begin{eqnarray}
N_S^{\pi^0/\eta} = N_{\text{POT}}\cdot f_{\pi^0/\eta} \cdot \text{BR}(\pi^0/\eta\to \gamma \gamma') \cdot \epsilon_{\text{cut\&acc.}}^{\pi^0/\eta}  \cdot \text{BR}(\gamma' \to e^+ e^-), \label{eqn:NS_MD}
\end{eqnarray}
where $N_{\text{POT}}$ is the total number of POT, equal to the number of POT per year times the number of operation years, $f_{\pi^0/\eta}$ is the fragmentation factor of the corresponding meson defined as the number of the meson produced for each POT, and $\epsilon_{\text{cut\&acc.}}^{\pi^0/\eta}$ is the average probability of the dark photons produced from $\pi^0/\eta$ decays to both pass the lepton kinematic cuts and decay inside the fiducial volume (``acc.'' stands for ``acceptance'').
\begin{table}[t]
\centering
\begin{tabular}{|c|c|c|c|}
\hline
$E_p$       & 600~MeV & 2~GeV  & 9.3~GeV \\ \hline
$f_{\pi^0}$ & 0.034   & 0.713  & 3.76    \\ \hline
$f_{\eta}$  & 0       & 0.0230 & 0.129  \\ \hline
\end{tabular}
\caption{Fragmentation factors of the $\pi^0$ and $\eta$ mesons, $f_{\pi^0}$ and $f_{\eta}$, for proton energies $E_p$ of 600~MeV, 2~GeV, and 9.3~GeV. The fragmentation factors are defined as the number of mesons produced for each POT.}
\label{tab:fragmentation}
\end{table}
In Table~\ref{tab:fragmentation} we list the values of $f_{\pi^0}$ and $f_{\eta}$ for proton energies $E_p=600$~MeV, 2~GeV, and 9.3~GeV, respectively.
We have derived these values by performing detailed simulations with the tool GEANT4~\cite{GEANT4:2002zbu}.
In the simulations, we have assumed a copper target of size $800\times800\times 1200$~mm$^3$, with the proton beam direction being 1200 mm long passing through the target.
We have used the physics list \textit{QGSP\_BIC\_HP} in the GEANT4 simulation, and counted the primary $\pi^0$ and $\eta$ mesons produced inside the whole copper target.
We have checked and found that the exact choice of the physics list of GEANT4 mainly affects the production rates of the considered mesons by up to about 50\%.
This uncertainty of 50\% would alter the final sensitivity reach to the dark-photon kinetic mixing to a minor extent, and therefore we do not take it into account in the computation but only comment briefly here.
For the computation of BR$(\pi^0/\eta\to\gamma\gamma')$ and BR$(\gamma'\to e^+ e^-)$, we refer to Sec.~\ref{sec:model}.
Moreover, the kinematics of the initial pions and eta mesons are provided by the GEANT4 simulations.

In order to determine $\epsilon_{\text{cut\&acc.}}^{\pi^0/\eta}$, we apply the technique of Monte-Carlo simulation.
Scanning over decades of values of $m_{\gamma'}$ and $\epsilon$, we simulate the $\pi^0$ and $\eta$ meson decays, separately, into an SM photon and a dark photon, followed subsequently by the dark-photon decays into an electron-positron pair.
For the $i^\text{th}$ simulated signal event, we first determine if the electron and positron pass the lepton cuts, and if so, compute the probability of the simulated dark photon in the event to decay inside the fiducial volume:
\begin{eqnarray}
\epsilon_{\text{cut\&acc.}}^{i^\text{th}\,\gamma', \pi^0/\eta} =
\begin{cases}
0, \text{ if failing the lepton cuts,}\\
\epsilon_{\text{acc.}}^{i^\text{th} \gamma', \pi^0/\eta}, \text{ if passing the lepton cuts.}
\end{cases}\label{eqn:cut_acc_individual}
\end{eqnarray}
The individual decay probability, $\epsilon_{\text{acc.}}^{i^\text{th}\,\gamma',\pi^0/\eta}$, is calculated as follows,
\begin{eqnarray}
    \epsilon_{\text{acc.}}^{i^\text{th}\,\gamma',\pi^0/\eta} = 
    \begin{cases}
    0, \text{ if }p_z^{i^{\text{th}}\,\gamma',\pi^0/\eta} \leq 0,\\
    0, \text{ if }\frac{p_T^{i^{\text{th}}\,\gamma',\pi^0/\eta}}{p_z^{i^{\text{th}}\,\gamma',\pi^0/\eta}} \geq \frac{r}{d},\\ 
    e^{-d/l_d^z}\cdot(1 - e^{-l/l_d^z}),
    \end{cases}\label{eqn:acc}
\end{eqnarray}
where $p_{T/z}^{i^{\text{th}}\,\gamma',\pi^0/\eta}$ is the momentum magnitude in the transverse or longitudinal direction of the $i^{\text{th}}$ simulated dark photon from either $\pi^0$ or $\eta$ meson decay, $r$ ($l$) is the radius (length) of the proposed detector, and $d=10$~m is the distance from the beam dump to the near end of the detector's fiducial volume.
For the CiADS-BDE, we have taken $r=0.1$ or 1~m, and $l=1$~m.
$l_d^z=\beta_i^z \gamma_i c\tau_{\gamma'}$ is the decay length of the $i^{\text{th}}$ simulated dark photon along the $z$-direction, with $\beta_i^z$ and $\gamma_i$ denoting respectively its speed in the $z$-direction and its boost factor, as well as $c\tau_{\gamma'}$ labeling the proper decay length of the dark photon.
Note that when simulating the meson decays, we assume that the negative $z$-direction is where the incoming proton beam comes from and that the beam dump is point-like\footnote{This approximation is justified because the considered mesons, $\pi^0$ and $\eta$, both decay promptly at the beam dump and the dump itself is only 1.2~m long in contrast to the 10 m distance to the detector behind it.} and located at the origin of the coordinate system, without loss of generality.
We thus evaluate $\epsilon_{\text{cut\&acc.}}^{\pi^0/\eta}$ by averaging over the simulated events,
\begin{eqnarray}
	\epsilon_{\text{cut\&acc.}}^{\pi^0/\eta} =\frac{1}{N_{\text{sim.}}}\sum_{i=1}^{N_{\text{sim.}}} \epsilon_{\text{cut\&acc.}}^{i^\text{th}\,\gamma',\pi^0/\eta} \,, \label{eqn:cut_acc_average}
\end{eqnarray}
where $N_\text{sim.}$ is the number of the simulated events.

\subsection{Proton bremsstrahlung}\label{subsec:NS_protonbrems}

We compute the signal-event number in the proton-bremsstrahlung process with
\begin{eqnarray}
N_S^{\text{PB}}&=& N_\text{POT}  \int d E_k \int_{\tan \theta_k < r/d }  d\cos\theta_{k}  \frac{1}{\sigma_{pCu}} \frac{d^2 \sigma_{\rm PB}}{d E_{k} d\cos\theta_{k}} \nonumber \\
&&\cdot  \, \epsilon^{\text{PB}}_{\text{cut\&acc.}}(\cos{\theta_k},E_k) \cdot \text{BR}(\gamma'\to e^+ e^-).\label{eqn:NS_PB}
\end{eqnarray}
Here, the explicit expressions for the upper and lower limits of the $E_k$ integral are provided in Ref.~\cite{Du:2022hms}, and $\sigma_{pCu}$ is the total cross section between the beam proton and the copper dump (see Sec.~\ref{subsubsec:model_pb}).
Further, $\epsilon^{\text{PB}}_{\text{cut\&acc.}}(\cos{\theta_k},E_k)$ is the equivalent of $\epsilon_{\text{cut\&acc.}}^{i^\text{th}\gamma', \pi^0/\eta}$  of the meson-decay case, and its computation follows closely Eq.~\eqref{eqn:cut_acc_individual} and Eq.~\eqref{eqn:acc}.
For the detailed calculation procedure of the production cross section of the dark photon in proton bremsstrahlung, see Sec.~\ref{subsubsec:model_pb}.

\section{Numerical results}\label{sec:results}

\begin{figure}[t]
	\centering
	\includegraphics[width=0.495\textwidth]{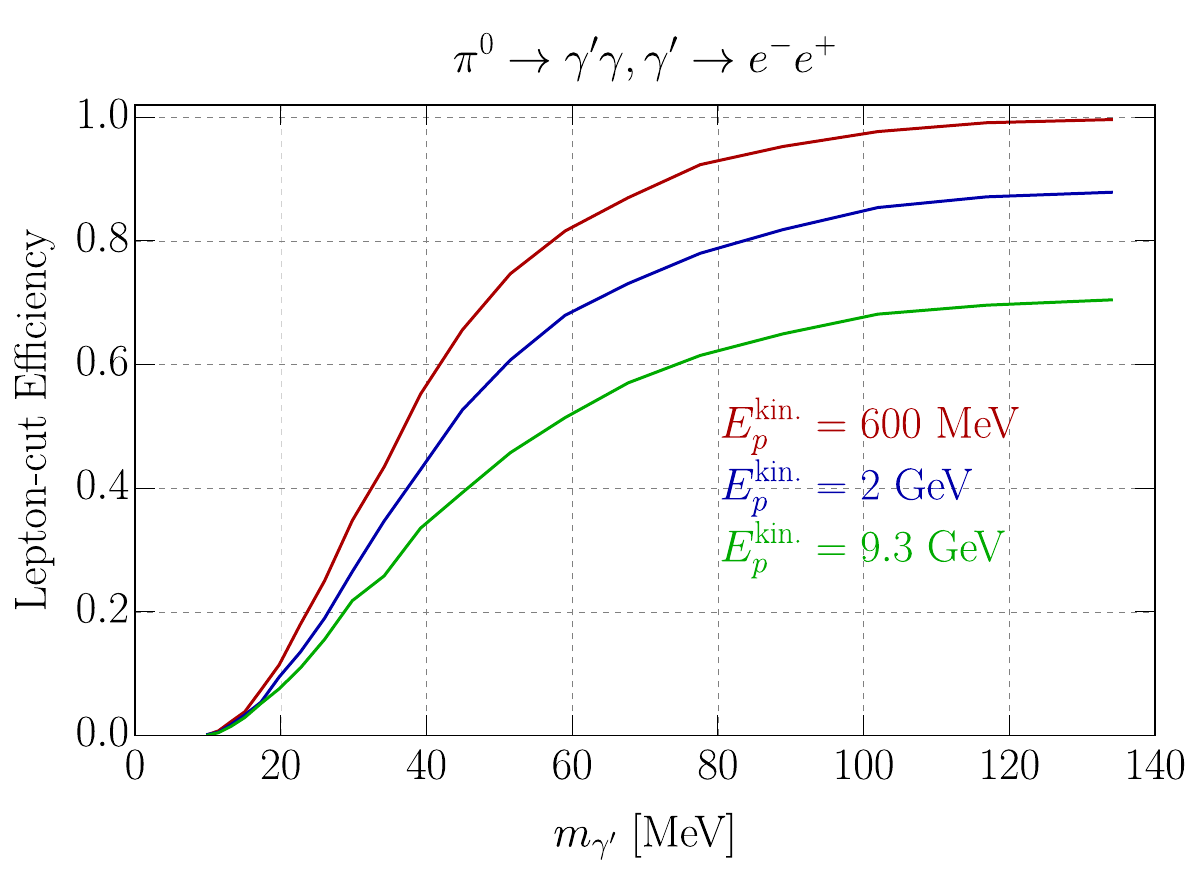}
	\includegraphics[width=0.495\textwidth]{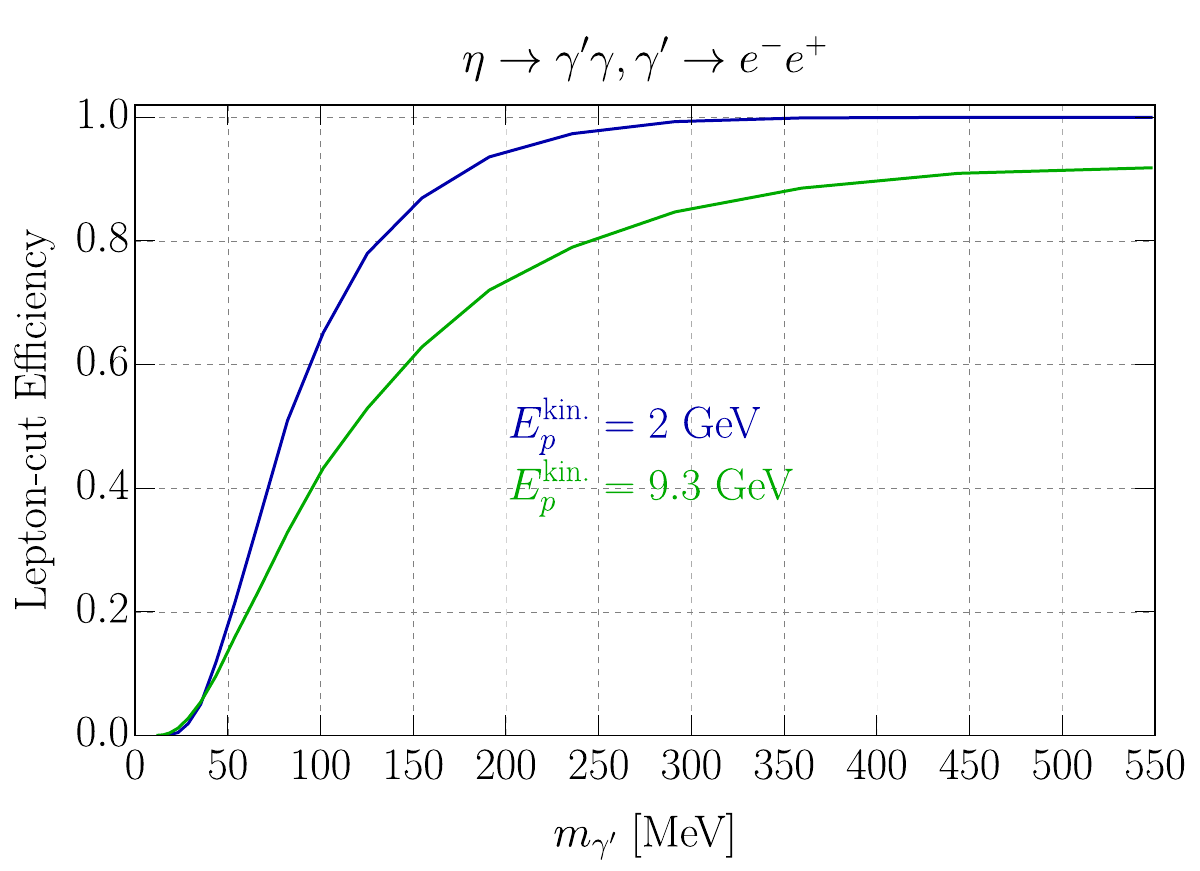}
	\includegraphics[width=0.495\textwidth]{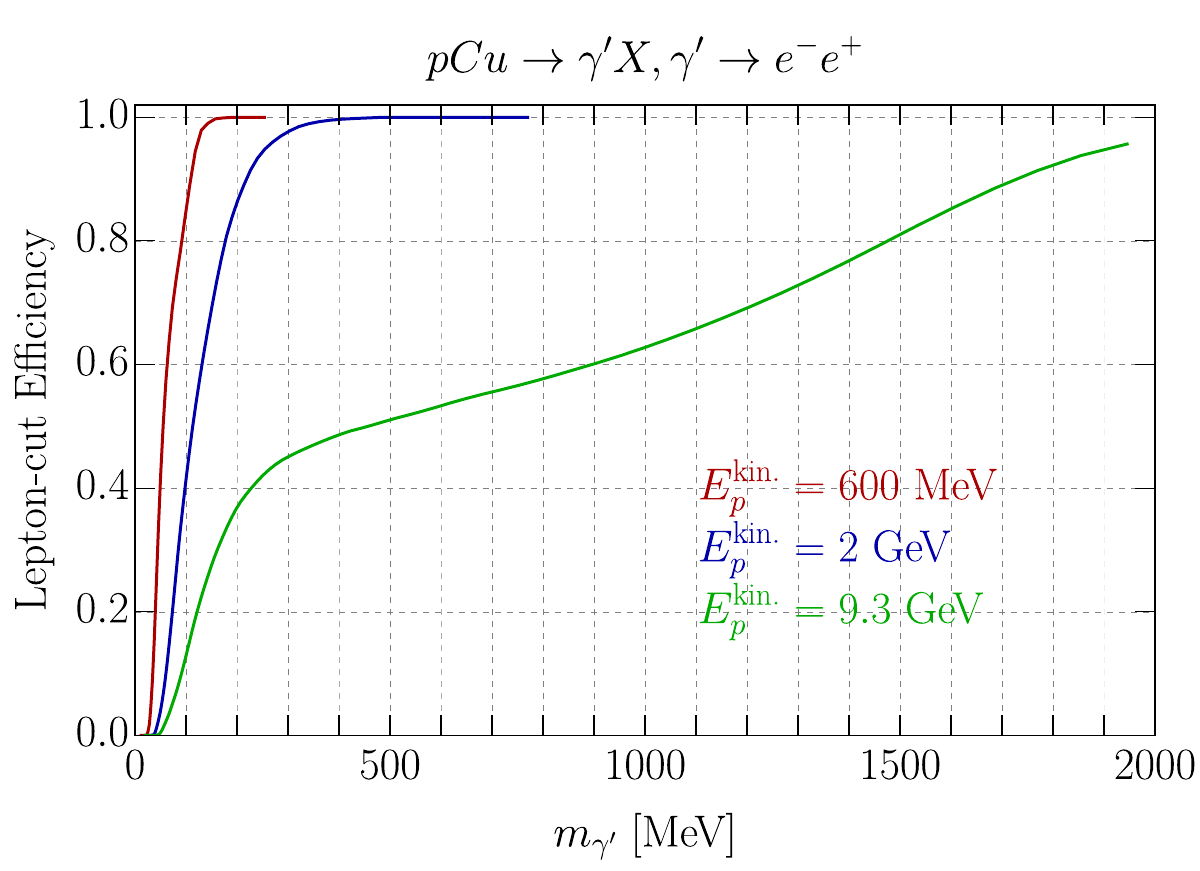}
	\caption{Lepton-cut efficiencies as functions of the dark-photon mass, for dark photons from $\pi^0$ decays (upper-left panel), $\eta$ decays (upper-right panel), and proton bremsstrahlung (lower panel), respectively.
 }
 \label{fig:leptoncut_efficiency}
\end{figure}

We present numerical results in this section.
We first show in Fig.~\ref{fig:leptoncut_efficiency} three plots of the lepton-cut efficiencies as functions of the dark-photon mass, corresponding to dark photons produced in $\pi^0$ decays, $\eta$ decays, and proton bremsstrahlung, respectively.
In these plots, we use the colors red, blue, and green, to denote the proton kinetic energies of 600~MeV, 2~GeV, and 9.3~GeV, respectively.
The minimal dark-photon mass is taken to be 10~MeV.
Note that here only the kinematic cuts on the leptons (the energy of both the electron and the positron should be larger than 17 MeV and their opening angle should be wider than $30^\circ$) are included, while the requirement on the fiducial volume is not.
In the plot for $\eta$ decays there is no result for proton energy of 600 MeV, because with this setup, no $\eta$ meson is produced.
We observe that in general with a more energetic proton beam or for smaller mass values, the lepton-cut efficiencies are reduced, mainly because the leptons have a larger boost thus tending to fail the opening-angle requirement.
For high dark-photon masses, the lepton-cut efficiencies reach 100\%.

\begin{figure}[t]
	\centering
	\includegraphics[width=0.495\textwidth]{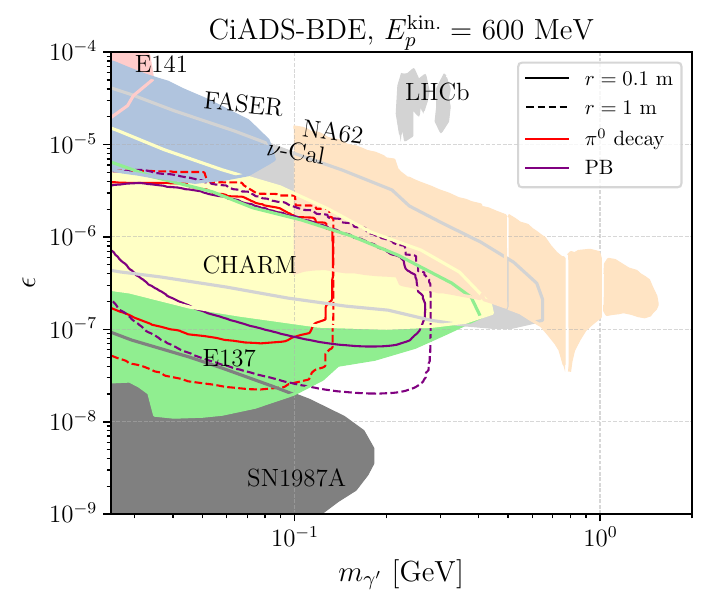}
	\includegraphics[width=0.495\textwidth]{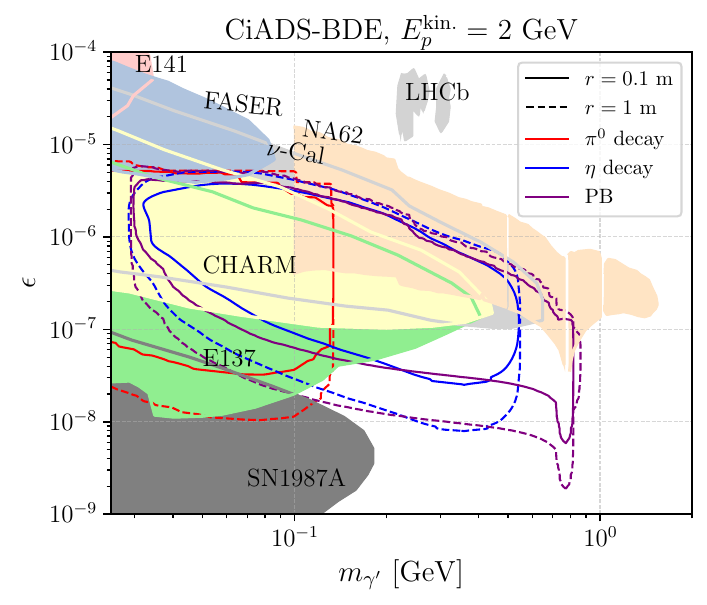}
	\includegraphics[width=0.495\textwidth]{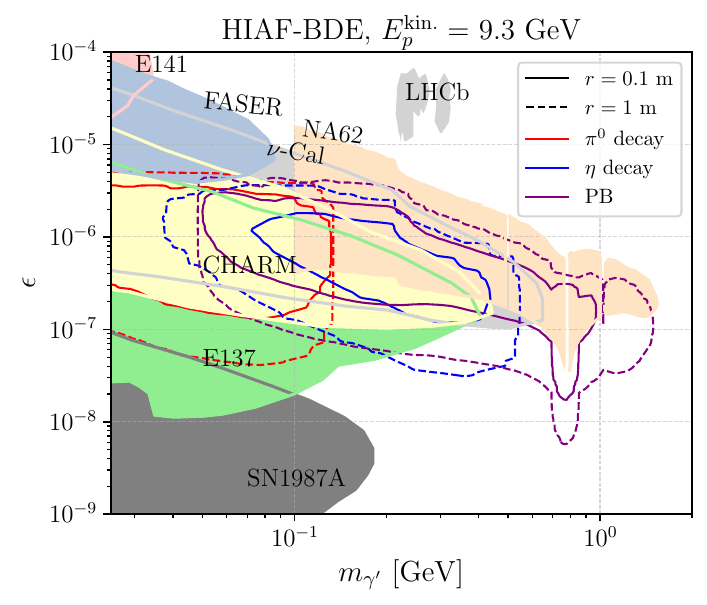}
	\caption{Sensitivity reach at $95\%$ C.L.~of CiADS-BDE or HIAF-BDE shown in the plane $\epsilon$ vs.~$m_{\gamma'}$, for dark photons produced in $\pi^0$ decays (red), $\eta$ decays (blue), and PB (purple).
    The three plots are for proton beam energies of 600~MeV (CiADS-BDE), 2~GeV (CiADS-BDE), and 9.3~GeV (HIAF-BDE), respectively, and in each plot both results for radius $r= 0.1$~m (solid) and 1~m (dashed) are displayed, together with the existing constraints obtained from LHCb~\cite{LHCb:2019vmc}, FASER~\cite{FASER:2023tle}, beam-dump experiments E141~\cite{Riordan:1987aw}, E137~\cite{Bjorken:1988as,Batell:2014mga,Marsicano:2018krp}, $\nu$-Cal~\cite{Blumlein:2011mv,Blumlein:2013cua}, CHARM~\cite{Gninenko:2012eq}, and NA62~\cite{NA62:2025yzs}, as well as supernovae~\cite{Chang:2016ntp}.
            }
 \label{fig:results_bdg}
\end{figure}

We show in Fig.~\ref{fig:results_bdg} the sensitivity reach at $95\%$ C.L.~of CiADS-BDE (upper plots) and HIAF-BDE (lower plot), to the kinetic mixing of the dark photon as functions of $m_{\gamma'}$.
The contour curves correspond to 3 signal events.
In these plots, the red, blue, and purple contour curves are for dark-photon production in $\pi^0$ decays, $\eta$ decays, and PB, respectively, and the solid (dashed) curves are for the detector radius of 0.1~m (1~m).
Leading existing bounds in the relevant mass range are extracted from LHCb~\cite{LHCb:2019vmc}, FASER~\cite{FASER:2023tle}, beam-dump experiments E141~\cite{Riordan:1987aw}, E137~\cite{Bjorken:1988as,Batell:2014mga,Marsicano:2018krp}, $\nu$-Cal~\cite{Blumlein:2011mv,Blumlein:2013cua}, CHARM~\cite{Gninenko:2012eq}, and NA62~\cite{NA62:2025yzs}, as well as supernovae observations~\cite{Chang:2016ntp}.
We find that for the CiADS-BDE only in the case of the proton-beam energy of 2~GeV, large new parameter regions beyond the present bounds can be probed, with dark photons produced in $\eta$ decays and proton bremsstrahlung.
Specifically, for $m_{\gamma'}$ roughly between 120~MeV and 800~MeV, the CiADS-BDE with $r=1$~m can exclude values of $\epsilon$ up to about one-to-two orders of magnitude lower than the presently strongest bounds that were obtained at the E137, $\nu$-Cal, and NA62 experiments.
In particular, compared to the $\eta$ decays, the PB process can be sensitive to heavier dark photons with smaller kinetic mixing.
For $r=0.1$~m, although the $\eta$ decays can only exclude a smaller corner extruding the present bounds, the PB process can extend the mass reach largely.
For the HIAF-BDE with $E_p^{\text{kin.}}=9.3$~GeV, despite the much smaller duty factor at the facility, the larger proton-beam energy at the experiment allows to probe dark-photon masses even beyond 1~GeV, with $r=1$~m.
For $m_{\gamma'}$ between approximately 300~MeV and 1200~MeV, $\epsilon$ of the order of $\mathcal{O}(10^{-8}\text{--}10^{-7})$ can be excluded.
We also comment that there are islands and dips of sensitivity in some of these plots and these particular features arise because of the strong resonance contributions to the dark-photon decay widths (see (Eq.~\eqref{eq:PB_formfactor} and Fig.~\ref{fig:dp_decay})).

\section{Conclusions}\label{sec:conclusions}

Recent years have witnessed quick development of LLP searches on both theoretical and experimental sides in the high-energy-physics community.
This is due to not only the absence of concrete discovery of new physics from searches for heavy fundamental particles at the LHC, but also the strong theoretical motivation of LLPs predicted in a wide range of BSM theories.
In particular, LLPs at the GeV- or even lighter scales have attracted much attention, partly because of the difficulty of searching for such light particles at the LHC main detectors.
Multiple supplementary experiments as far detectors at the LHC, and beam-dump experiments have been either proposed or operated, in order to fill this gap.
These proposals have been shown to allow for testing uncharted territories of various LLP scenarios.

In this work, we have proposed a beam-dump experiment at CiADS located in Huizhou, Guangdong, China, called CiADS-BDE, where a proton beam with kinetic energy around 1~GeV is planned to impinge on a high-power beam dump made of oxygen-free copper.
Such low-energy proton collisions could lead to both primary and secondary production of light LLPs that tend to travel in the forward direction.
The CiADS-BDE should place a detector 10~m behind the dump, and the LLPs could decay inside this detector.
The space between the detector and the dump is available for implementation of shielding material and veto components, largely suppressing the background events.
Moreover, the detector is supposed to be of a cylindrical shape, consisting primarily of liquid scintillator.
It has a length of 1~m and a radius of either 0.1~m or 1~m.
We assume a 5-year beam commissioning time.

Given the projected $\mathcal{O}(10^{23})$ POT per year at the CiADS, we expect strong sensitivity reach to sub-GeV- to GeV-scale LLPs at this intensity frontier.
In this work, we have chosen the dark photon as a representative model in order to evaluate the search potential of the CiADS-BDE for light LLPs.
We focus on the signature of electron-positron-pair events, imposing kinematic cuts on the final-state leptons for suppressing background events and taking into account the dark photon's decay probability inside the fiducial volume of the proposed detector.
For the considered experimental setup, we have included in the computation the dominant production channels of the dark photon, i.e.~(1) rare decays of neutral mesons ($\pi^0$ and $\eta$) into an SM photon and a dark photon, and (2) proton bremsstrahlung.
We note that for the considered signature, we have estimated the background levels with dedicated simulations, with kinematic cuts on the lepton energy and opening angle identical to those applied to the signal events.

We have performed Monte Carlo simulations to generate spectra of mesons, as well as the dark photons and the first-generation charged leptons produced therefrom, and computed the signal-event rates at the proposed experiments.
For the PB process, we have calculated the proton-copper differential scattering cross sections into a dark photon plus anything and the signal-event rates with formulas given in the literature.
Our numerical results show that the lepton-cut efficiencies increase with heavier dark photons and smaller proton-beam energies, where the fiducial-volume requirement is not included.
The final sensitivity results show that the CiADS-BDE can probe new parameter space currently unexcluded, in the dark-photon mass range roughly between 0.12~GeV and 0.8~GeV and the kinetic-mixing parameter $\epsilon$ of the order of magnitude $\mathcal{O}(10^{-9}\text{--}10^{-8})$, for proton-beam kinetic energy of 2~GeV and the detector radius of 1~m.
Besides the CiADS, we have considered the possibility of setting up a similar experiment at HIAF located near the site of CiADS, called HIAF-BDE, to be operated for 5~years.
With a proton beam of higher energy but a lower duty factor, we find that the HIAF-BDE cannot exclude lower values of $\epsilon$ but is sensitive to heavier dark photons, now beyond 1~GeV.

In addition to the beam dump, the CiADS research reactor~\cite{LI2025103197} intrinsically provides dump-like conditions, rendering itself also suitable or complementary for such experiments.
Meanwhile, we estimate that the cost of installing such a beam-dump experiment should be relatively low, since the already planned proton beam can be utilized and the requirement of the detector system is minimal.
Moreover, the dark photon is only one benchmark model we choose in order to illustrate the potential of the proposed BDEs in LLP searches.
In principle, further physics goals can be pursued at these experiments.
For instance, numerous additional LLP scenarios can be explored.
Furthermore, neutrino physics can be probed including neutrino oscillation and unitarity violation of the active-neutrino mixing matrix.
Considering the relatively low cost, the rich list of potential physics applications, and that  the CiADS program is currently under construction and HIAF has started operation, we believe that it is good timing now to plan, design, and install the proposed experiments, and to perform further phenomenological studies on the setups.

\acknowledgments

We thank Hanjie Cai and Guodong Shen for useful discussions.
We acknowledge support by the National Natural Science Foundation of China under Grants No.~12475106 and No.~12505120, the Fundamental Research Funds for the Central Universities under Grant No.~JZ2025HGTG0252.
LW Chen is supported by the National Natural Science Foundation of China (Grant No.~12105327) and the Guangdong Basic and Applied Basic Research Foundation (Grant No.~2023B1515120067).

\appendix

\bibliographystyle{JHEP}
\bibliography{ref_list}

\end{document}